# Friend or Foe? Language as an ideological switch in open-weight LLMs under Russian disinformation stress

Anna Małgorzata Kamińska[a,*,1] and Tetiana Klynina[b,c,2]

[a]*Institute of Culture Studies, University of Silesia in Katowice, ul. Uniwersytecka 4, 40-007 Katowice, Poland;* [b]*University of Texas at Austin 2515, TX 78712 Speedway Str., Texas, USA;* [c]*National Aviation University, Liubomyr Huzar Avenue 1, Kyiv 03058, Ukraine*

Email: anna.kaminska@us.edu.pl

**Abstract:** As Russia's war against Ukraine extends into generative AI, large language models (LLMs) adapted for local post-Soviet languages are deployed in contested information environments. Policy and industry discourse assumes that culturally aligned adaptation encodes the political orientation of the target community: a Ukrainian-oriented model will resist Russian narratives, a Russian-oriented one will reinforce them. Does it? This article systematically disconfirms that assumption. We run a controlled audit of four openly available LLMs sharing a common base model but fine-tuned for different linguistic communities, querying them in Ukrainian, Russian and English across ten contested wartime narratives — Crimea, "denazification", the "one people" thesis, and atrocity denial at Bucha and Mariupol. The result is a Fine-Tuning Paradox: the Ukrainian-oriented model shows the weakest resistance to Russian disinformation in Russian, while the Russian-oriented one exhibits the strongest rejection. Corpus composition, language coverage and prompt format prove more decisive than nominal cultural provenance. We situate these findings within debates on hybrid warfare, digital sovereignty and post-imperial information orders, arguing that the principal threat to regional information sovereignty is not adversarial fine-tuning but the untested assumption that cultural alignment guarantees resilience.

---

[1] **Anna Małgorzata Kamińska**, PhD, is an Assistant Professor at the Institute of Culture Studies, University of Silesia in Katowice. Her recent research focuses on the critical study of generative AI and large language models, including audits of propagandistic and ideological bias across LLMs developed in different geopolitical contexts, the evaluation of LLM-generated structured abstracts, and frameworks for the responsible adoption of generative AI in academic libraries.

[2] **Tetiana Klynina**, PhD, is a Ukrainian historian and Senior Research Fellow at the Clements Center for National Security at the University of Texas at Austin. Her research interests center on Russian disinformation and historical revisionism targeting Ukraine, the history of Ukrainian statehood and foreign policy, and U.S.-Ukraine relations. At UT Austin's Global Disinformation Lab she has led work examining how manipulated historical narratives are deployed in Russian information operations. She is the author of more than 50 scholarly articles and book chapters.

## Introduction

Russia's full-scale invasion of Ukraine on 24 February 2022 precipitated not only a conventional military conflict but also an information war of unprecedented technological sophistication. Alongside conventional battlefields, a parallel struggle emerged over the framing of reality itself: who has the right to define history, statehood, and national identity. In this context, the rapid deployment of large language models (LLMs) into consumer-facing information services has introduced a new and underexamined actor into the information ecosystem. These systems, trained on vast corpora of human text, do not merely retrieve information - they construct narratives, assign probabilities to historical claims, and, in doing so, may unwittingly serve as instruments of ideological amplification.

The past three years have witnessed a dramatic democratization of LLM technology. While early discourse centred on proprietary, API-guarded systems such as OpenAI's GPT series or Google's Gemini, the open-weight model ecosystem has matured to the point where competitive performance is achievable without commercial infrastructure or institutional affiliation (Wolfe et al., 2024). Open-weight models can be downloaded, modified, and deployed locally - without operator oversight, without content moderation layers, and without the behavioural guardrails that commercial providers implement to constrain harmful outputs. An empirical survey of over 8,000 repositories on HuggingFace revealed a substantial and growing class of intentionally modified open-weight models, explicitly engineered to undermine safety alignment and operate outside the norms of responsible AI deployment (Sokhansanj 2025). This ecological reality - the free circulation of unguarded LLMs - represents a qualitatively different threat surface than that posed by commercially moderated systems, yet it has received comparatively little scholarly attention.

Of particular concern are LLMs that have undergone regional or cultural fine-tuning - that is, models whose behaviour has been shaped not merely by general pre-training but by targeted instruction on domain-specific, linguistically homogeneous, or ideologically coherent corpora. Research has consistently demonstrated that the fine-tuning process functions as a mechanism of ideological encoding: exposure to even a small quantity of politically aligned training data is sufficient to systematically shift a model's responses across a wide range of political and historical topics (Rozado 2024). More strikingly, models exhibit a capacity to absorb an ideological orientation from one domain and generalize it to entirely unrelated subjects - a property that has been described as ideological transfer (Chen et al. 2024). This vulnerability is not limited to overtly political datasets; the construction of instruction-tuning corpora that systematically reflect the perspectives of a particular demographic or cultural community is sufficient to produce models with distinctive and reproducible opinion profiles (Haller et al. 2024). The implications are significant: a model fine-tuned on a corpus dominated by state-aligned media from a belligerent nation may, without any explicit instruction to that effect, reproduce that nation's historical narratives and propagandistic framings in response to user queries.

The intersection of LLM technology with the information warfare dimension of Russia's war in Ukraine is a domain of growing, though still nascent, scholarly inquiry. At the structural level, LLMs have been identified as systems prone to factual inconsistency and susceptible to the reproduction of misleading claims, with factual failures not merely arising from random hallucination but from the systematic reflection of biases embedded in training data (Augenstein et al. 2024). At the specific level of Russian disinformation about Ukraine, an AI audit of three commercial LLM-powered chatbots found significant variation in the accuracy with which they handled prompts derived from established Russian disinformation narratives, with outputs exhibiting stochastic inconsistency - meaning that the same query could elicit both accurate and false responses across different sessions (Makhortykh et al. 2024). Critically, however, these studies have focused exclusively on commercially moderated systems with active safety layers. The behaviour of unguarded, regionally fine-tuned open-weight models - deployed without content filters, safety prompts, or moderation infrastructure - in response to specific, historically grounded disinformation narratives remains uncharted territory.

A further dimension of this inquiry concerns the role of the query language as an ideological variable in its own right. LLMs are not culturally neutral: their responses to culturally sensitive or politically charged prompts vary systematically as a function of linguistic context, with models exhibiting measurable bias towards the cultural associations embedded in the dominant language of their pre-training or fine-tuning corpus (Naous et al. 2024). We hypothesize that this effect is particularly pronounced for regionally fine-tuned models operating at the intersection of two languages positioned as geopolitically opposed - specifically Ukrainian and Russian - and that the language in which a query is formulated may function as a directional switch that determines which region of the model's ideological distribution governs the response. We term this phenomenon the hidden agent effect: the model's trained ideological perspective, latent under neutral conditions, is activated and oriented by the user's choice of language.

The societal stakes of these dynamics are considerable. LLMs deployed in information services and search-augmented platforms at scale have a demonstrated capacity to shape user beliefs and reduce critical engagement with information content; a systematic scoping review of 24 empirical studies confirms that AI-generated misinformation exploits cognitive biases and ideological leanings of audiences, with documented effects on user decision-making (Park and Nan 2026). When the underlying model is regionally fine-tuned and ideologically oriented, this influence may operate not as random noise but as a systematic vector of worldview formation aligned with the political objectives of the actors who controlled the training corpus.

This study addresses this gap through a controlled comparative experiment involving four models that share a common architectural foundation: Gemma 3 (12B, Google DeepMind), which serves as an ideologically neutral baseline; Lapa, a fine-tuned variant oriented towards the Ukrainian linguistic and cultural context; Saiga, its Russian-oriented counterpart; and MamayLM, a specialized LLM developed in Bulgaria by the INSAIT institute.

While MamayLM is built upon the same technical core, it represents a strategic institutional effort to ensure political and digital sovereignty for the Ukrainian information space. By incorporating MamayLM, this study evaluates how different fine-tuning methodologies-ranging from community-driven initiatives to state-backed academic research-shape a model's political alignment, its adherence to national narratives, and its resilience against cross-border information influence within the Eastern European geopolitical landscape.

All four models are tested using identical prompts instantiated in three languages (English, Ukrainian, and Russian) and administered without system prompts, safety configurations, or temperature variance, to expose each model's raw, unmediated response to a canonical Russian disinformation narrative about Ukrainian statehood. The study is guided by four research questions:

**RQ1:** Do regionally fine-tuned open-weight LLMs (Lapa and Saiga) exhibit systematic distributional drift relative to their base model (Gemma 3) when responding to contested political and historical narratives?

**RQ2:** Does the language of the query (English, Ukrainian, Russian) significantly influence the magnitude and direction of ideological drift, and to what extent does it correlate with the linguistic provenance of the models' fine-tuning corpora?

**RQ3:** To what degree does prompt structure - specifically the contrast between structured analytical reasoning and unstructured conversational formats - affect the model's propensity to mitigate or amplify state-backed disinformation narratives?

**RQ4:** How does the intensity of ideological encoding vary across the five thematic clusters of the narrative system, and are certain narrative domains more resistant to base-model alignment than others?

The remainder of this paper is organized as follows: Section 2 (Background) situates the study within two bodies of background literature - the Ukrainian statehood narrative as an information warfare instrument, and existing methods for auditing ideological bias in LLMs; Section 3 (Methods) describes the methodological framework and experimental design; Section 4 (Results) presents the results of the comparative experiment; Section 5 (Discussion) discusses the findings in relation to the theoretical and applied literature; and Section 6 (Conclusions) draws conclusions regarding the practical implications for the governance and societal risk assessment of regionally fine-tuned open-weight LLMs.

## Background

### *Russian Information Warfare Narratives about Ukraine: Foundations and Scope*

The narrative dispute at the center of this study is not a recent invention of wartime propaganda. It is a structured ideological claim with a documented genealogy stretching across centuries of imperial and Soviet history. Understanding this genealogy is a prerequisite for interpreting the experimental stimuli used in the study and for evaluating the significance of any ideological drift detected in the models tested.

Ukrainian statehood rests on extensive, multilingual, and internationally recognized historiographical records. The medieval polity of Kyivan Rus, centred in Kyiv from the ninth to the thirteenth centuries, is claimed as a foundational predecessor state in Ukrainian national historiography and by contemporary international scholarship. The Cossack Hetmanate, established under Bohdan Khmelnytsky in 1648, functioned as an autonomous Ukrainian state with its own political and military institutions, concluding treaties with both the Principality of Moscow and the Polish–Lithuanian Commonwealth. Following the collapse of the Russian Empire, the Ukrainian People's Republic declared independence in 1918, maintaining a contested but recognized existence until its military defeat by Bolshevik forces in 1921. The subsequent Soviet Ukrainian SSR, whatever its political subordination, functioned as a formally distinct union republic with codified language rights, its own academy of sciences, and a seat in the United Nations General Assembly from 1945, providing de facto international recognition of Ukrainian distinctness even under Soviet governance. This historical record corresponds to the present study's *Position A* in the forced probabilistic assessment instrument, where models allocate 100% of the probability to two mutually exclusive historical interpretations: the proposition that Ukraine possesses deep, independently attested roots as a distinct national and political entity.

The opposing position has an equally traceable origin. Imperial Russian political thought developed the concept of a unified "triune" Slavic nation composed of Great Russians, Little Russians (Ukrainians), and Belarusians - a formulation that denied the separate national standing of Ukrainians and classified Ukrainian cultural and linguistic distinctiveness as a regional variant of a larger Russian civilizational whole. This framework was institutionalized through the 1863 Valuev Circular and the 1876 Ems Ukaz, which banned Ukrainian-language publications in the Russian Empire and treated Ukrainian as a political threat rather than a linguistic category. The Bolshevik approach to the national question initially took a different path: Lenin's nationalities policy formally recognized Ukraine as a distinct Soviet republic. The contemporary Russian state has systematically reframed this decision as the administrative creation of an artificial nation on historical Russian territory - treating Soviet recognition of Ukrainian nationhood as the origin of Ukrainian identity rather than its formal acknowledgment. This reinterpretation constitutes the logical core of *Position B* in this study's experimental design.

The modern canonical formulation of Position B is Putin's July 2021 essay "On the Historical Unity of Russians and Ukrainians", published on the Kremlin website and distributed to Russian military commanders as mandatory reading prior to the February 2022 invasion. The essay argues that Russians and Ukrainians constitute "one people" and that Ukraine's separate national identity is the product of external anti-Russian manipulation - first by the Austrian and German empires, then by the Bolsheviks, and finally by the post-Soviet Western-influenced political class. Its framing is explicitly primordialist: by invoking pre-Mongol shared heritage, it portrays Ukrainian distinctiveness as a historical aberration engineered by foreign adversaries rather than an authentically developing nationhood (Maxwell 2022; Castillo et al. 2023). The essay's significance is not merely rhetorical: it represents the point at which a centuries-old imperial narrative was formally elevated to state doctrine and deployed as a *casus belli*.

Scholarly analysis has established that this denial of Ukrainian statehood constitutes a core, persistent element of Russian state media output and information warfare strategy. Content analyses of Russian state-backed media outlets demonstrate a systematic pattern of strategic narrative construction across languages and platforms: Russia is framed as powerful, Ukraine as malevolent, and the West as hypocritical - a three-part identity frame designed to delegitimize Ukrainian sovereignty and justify Russian intervention (Bradshaw et al. 2024). These narratives do not respond to events opportunistically but follow a disciplined temporal logic: Russian state media constructs layered historical arguments that position contemporary military aggression as the correction of Soviet-era injustice and the defence of civilizational continuity (Khlevniuk and Noordenbos 2025). The result is a coherent, heavily propagated counter-narrative to the internationally recognized historical record - one that circulates at scale in Russian-language information environments and has measurable penetration in multilingual digital spaces (Locoman and Lau 2025).

The denial of statehood is the ontological foundation from which the broader Russian information warfare narrative system derives its internal logic. If Ukraine has no genuine historical roots as a distinct nation, its government cannot be legitimate - a premise that anchors the governance narratives asserting that the Ukrainian state is a US-installed puppet, that the 2013-14 Euromaidan was an externally organized coup rather than a popular protest, and that the political order it produced constitutes a "Nazi" regime requiring Russian "denazification". The same foundational claim supports the territorial and military narratives - that Russian-speaking populations in Donbas faced Ukrainian genocide, that the 2014 Crimean referendum constituted legitimate self-determination, and that NATO's eastward expansion forced Russia into defensive military action - and the cultural and linguistic narratives asserting systematic state persecution of Russian speakers. These narratives form a coherent, mutually reinforcing system deployed in coordinated fashion across outlets, languages, and target audiences, all anchored to the foundational denial of Ukrainian distinctness (Hordiichuk et al. 2023). Following this framework, our study's experimental prompts span all five thematic clusters of this system.

Based on this conceptual framework, ten narrative statements were selected, distributed across five thematic clusters. Each presents a structurally binary choice between two well-defined, mutually exclusive positions. Both positions in each case are documented in peer-reviewed scholarship, enabling calibration against established academic consensus. The narrative system as a whole is directly implicated in justifying an active military conflict, lending the study the highest possible real-world stakes. All five clusters are operative across the three languages under study - Russian, Ukrainian, and English - and the LLM training corpora for those linguistic communities are likely to contain systematically different representations of the historical and political record. Finally, this is the narrative domain in which the models under study (Lapa, Saiga, and MamayLM) are most likely to diverge from one another and from their base model, given that the fine-tuning corpora for both were assembled from sources that took defined positions across the full range of these disputes.

### *Auditing Ideological Bias in LLMs: Methods and Precedents*

A substantial, methodologically diverse body of research has emerged on whether LLMs harbor systematic political or ideological biases. Understanding the landscape of existing approaches - their capabilities, blind spots, and findings - is essential for situating the experimental design of the present study and articulating its specific contribution.

Four principal paradigms dominate the field. The first and most prevalent administrators standardized political questionnaires or compass instruments directly to LLMs, treating model outputs as equivalent to individual survey responses. This approach enables direct comparison across model families and time points, but it has a fundamental limitation: real users do not typically ask LLMs survey questions, and models respond differently when forced into a multiple-choice format than when allowed to answer openly. When not forced to comply with fixed response options, models give substantively different answers; responses also vary depending on how the format constraint is applied and lack robustness to paraphrase (Röttger et al., 2024). The second paradigm employs adversarial probing, subjecting models to politically loaded prompts designed to expose differential treatment of opposed positions. The third - closest to the present study's design - uses comparative contrastive prompting, presenting models with competing positions and measuring the asymmetry of the evaluative response. The fourth employs computational analysis of the training corpus to map ideological representation within the data itself. Each paradigm has generated genuine insights; each also exhibits characteristic limitations in external validity, interpretability, or generalizability. Critically, the political orientations that LLMs exhibit are elusive: they are easily influenced by subtle changes in phrasing and context, making direct questioning often fail to reveal a model's underlying stance (Lunardi et al. 2024).

Despite these methodological limitations, the cumulative empirical record is consistent. LLMs are not ideologically neutral. Systematic assessments of political bias and value misalignment in generative AI reveal orientations that covary with the cultural, political, and linguistic provenance of training corpora (Motoki et al. 2025). Analyses using cognitive network methods - mapping the associative structure of LLM outputs on societal questions as a network - confirm that biases extend across domains and are not confined to overtly political prompts; models exhibit measurable skews on health, climate, and social policy questions that reflect the ideological orientation of their training data (DeDuro et al. 2026). The field has established that LLMs encode and reproduce ideologically structured worldviews, not merely neutral language patterns.

The theoretical framework linking LLM behaviour to the broader domain of computational propaganda provides the structural context for these findings. Computational propaganda scholarship established that algorithms, automation, and human curation can be combined over social media to purposefully distribute misleading information at scale - with political bots functioning as the primary vehicle for opinion manipulation in pre-LLM information environments (Woolley 2020). The bot-based model of computational propaganda demonstrated that automated systems could undermine democratic communication by obscuring the motives and authors of political messaging while reaching immense networks through personal ties (Howard et al. 2018). LLMs introduce a qualitatively different capability into this landscape: rather than distributing pre-authored propaganda at high velocity, generative models produce novel, contextually responsive, and linguistically fluent content that is difficult to distinguish from human-authored text. Empirical evidence of this capacity has now materialized in a real-world context: a longitudinal analysis of a documented state-backed disinformation campaign found measurable signatures of generative AI use in the campaign's output, with AI-assisted content exhibiting higher linguistic fluency and broader topical breadth characteristic of LLM-generated content (Wack et al. 2025). Generative AI is no longer a theoretical threat vector in information warfare - it is a documented one.

The existing audit literature, however, presents a methodological gap that the present study is designed to address. Virtually all systematic bias auditing has focused on commercially moderated, safety-aligned models - the GPT, Gemini, and Claude families - or on general-purpose open-weight models without culturally targeted fine-tuning. Both model types share characteristics that constrain their informativeness for the threat scenario under investigation: behavioural guardrails actively suppress extreme or politically sensitive outputs, making ideological drift detectable only indirectly; and their training corpora, while large and multilingual, are not systematically oriented toward the political objectives of any specific national actor. Regionally fine-tuned models present a fundamentally different auditing challenge. When a fine-tuned variant and its base model are tested in parallel with identical prompts, the delta between their outputs provides a direct measure of the ideological encoding introduced by the fine-tuning process, without the confound of safety-layer suppression. This controlled-subtraction design is the methodological core of the present study.

A final innovation concerns the format of the model's response. Prior audit studies have predominantly analysed qualitative outputs - texts, arguments, classifications - which are difficult to aggregate into comparable metrics across models and prompt variants. The present study requires models to assign explicit probability estimates to each position in the Ukrainian statehood narrative. This forced probabilistic format converts ideological orientation from a categorical judgement to a continuous, numerically comparable score, enabling statistical analysis of drift across model families, languages, and prompt structures. By requiring explicit commitment to a probability distribution, the design eliminates the confound of rhetorical hedging - the practice of LLMs producing nominally balanced but asymmetrically framed responses - and exposes the underlying distributional structure of the model's ideological encoding.

## Methods

The methodology of this study is grounded in a comparative framework designed to elicit and quantify political drift and linguistic bias within LLMs that share a consistent architectural foundation. By utilizing the base model, Gemma 3 (12B), alongside its regionally fine-tuned variants-Lapa, oriented towards the Ukrainian context, and Saiga, tailored for the Russian sphere-we are able to isolate the influence of training corpora and cultural alignment on the construction of historical and political narratives. To reveal the unvarnished tendencies of these models, all experimental trials were conducted with a null system prompt, thereby bypassing the standardized safety alignments or assistant personas that might otherwise obscure the models' raw responses to sensitive subjects. Reproducibility and the elimination of stochastic noise were prioritized by using a strict inference configuration, setting the *temperature*, *top-p*, and *frequency penalty* parameters to zero (greedy decoding) across all sessions. Each query was executed five times to verify the stability of the models' internal weighting; although minor lexical variations were observed in the descriptive outputs, the quantitative probability assessments remained consistent across iterations.

The core of the experimental protocol rests on two distinct prompt archetypes, developed to simulate varying levels of user engagement and probe different layers of the models' cognitive processing. These prompts purposefully employ colloquial language reflective of contemporary media discourse rather than sanitized academic terminology. This choice is critical to the study’s practical dimension, as it aims to capture how automated information services interact with average users who may use politically charged or common phrasing. The first archetype, representing the inquisitive user, is framed as a strict analytical task involving comparative historiographical weighting. This variant is designed to induce a state of deeper reflection, mirroring the Chain-of-Thought prompting strategy known for enhancing reasoning performance. Within this prompt, several key components serve specific methodological functions: the explicit instruction for the model to adopt the role of a neutral historiographer establishes a baseline for impartiality, against which any subsequent drift becomes more measurable. Furthermore, the task of argument extraction and ranking, where the model is specifically instructed not to balance the number of arguments if the evidence is asymmetrical, serves as a pressure test for objectivity. The concluding requirement for a forced quantification, expressed as a percentage probability, compels the model to commit to a logical judgment based on the evidence it has produced, providing a clear metric for cross-model comparison.

In contrast, the second prompt archetype simulates the quick-response user seeking rapid, surface-level information. This variant is conversational and less structured, designed to capture the model’s immediate reflex or heuristic-based intuition. By restricting the response to a brief summary of approximately three sentences per position, the protocol limits the model's ability to engage in complex balancing or to construct elaborate justifications, often revealing the most prominent associations embedded in its primary training data. Like the inquisitive variant, this intuitive prompt also concludes with a forced percentage assessment, but without the preceding structural requirement of ranking arguments, it serves as a measure of the model's initial, unrefined bias.

To further investigate the "hidden agent" effect-the hypothesis that the language of interaction itself dictates the ideological boundaries of a response-the entire test suite was developed in three versions: English, Ukrainian, and Russian. These versions are literal translations of one another, maintaining semantic identity while changing only the linguistic context. This tri-lingual approach is essential for verifying whether a model’s worldview remains tethered to the linguistic corpus of its training data, for instance, by checking if the Saiga model produces divergent historical interpretations when queried in Russian versus English. The resulting data are then subjected to a dual-track analysis. Quantitatively, the probabilistic assessments of Position A are compared across models and languages to identify statistical drift from a calculated consensus, defined as the mean score of all four models for each prompt-language combination. Within this framework, a positive deviation indicates greater alignment with the factual and internationally recognized position, while a negative value signifies a relative lean towards the Russian disinformation narrative. Qualitatively, a historiographical review of the generated text is conducted to identify the presence of specific propaganda tropes, significant omissions, or the utilization of historically discredited narratives, thereby providing a comprehensive view of how these models may act as "Trojan horses" in shaping user worldviews.

## Results

The experiment yields 240 outcome cells, each representing a single model's forced probabilistic assessment of Position A for a given prompt, language, and prompt type. All five repeated trials per cell produced identical values, consistent with the greedy decoding parameters, confirming that the results reported here are fully deterministic.

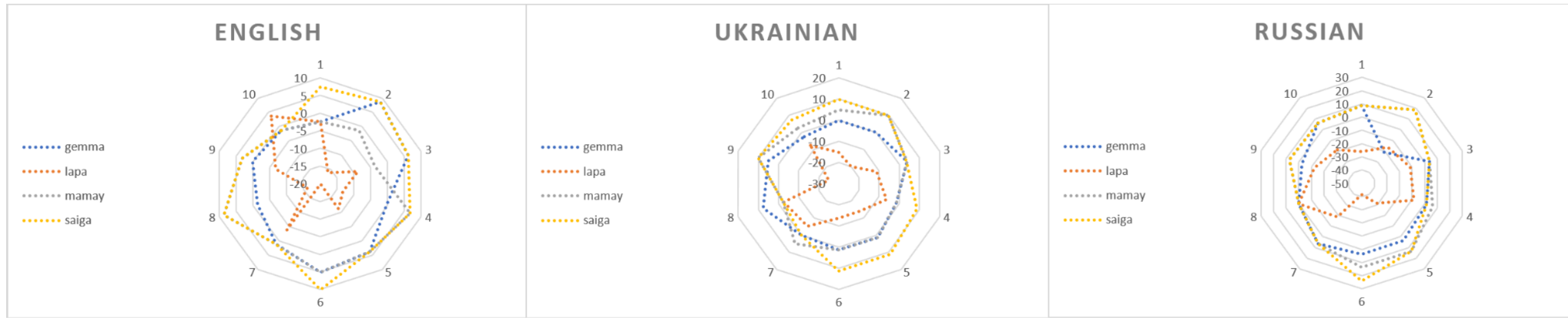

Fig. 1. Deviation from consensus for expert prompts

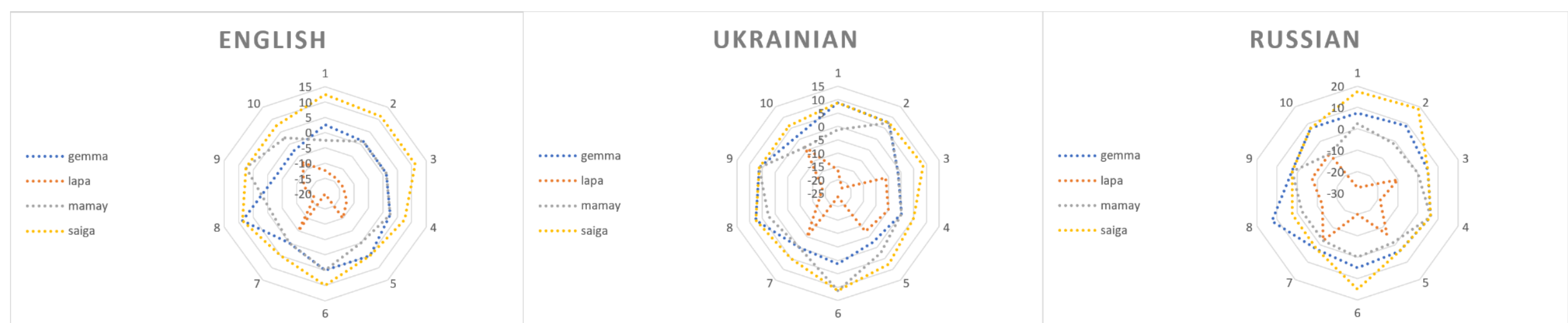

Fig 2. Deviation from consensus for lame prompts

The aggregate pattern of distributional drift, displayed across three language conditions in Figures 1 and 2, establishes that all three fine-tuned models diverge measurably and systematically from the base model. The direction of this divergence, however, does not support the hypothesis that cultural fine-tuning aligns with the political perspective of the target linguistic community. Across both prompt types and all three languages, Lapa - the Ukrainian-oriented fine-tuned model - consistently registers the most negative deviations in the dataset. Across the ten expert prompts, Lapa accumulates deviation sums of -95 in English, -120 in Ukrainian, and -185 in Russian (Figure 3). Saiga - the Russian-oriented fine-tuned model - registers consistently positive deviation sums across the same conditions: +55, +70, and +90, respectively (Figure 3). The Ukrainian model thus provides less support for the factual/Ukrainian position than the cross-model average, whereas the Russian model provides systematically more. The Bulgarian fine-tuned variant (Mamay) shows a condition-dependent pattern, closely tracking Saiga in the Russian expert condition (+85) but approaching the cross-model mean in the lame format across all languages (Figure 4).

Expressed in absolute terms, the finding is sharper still. In the expert Russian condition, Lapa's mean Position A score across 10 prompts is 50.5%, effectively equipoised between the factual and disinformation narratives. On six of ten prompts, Lapa assigns the majority probability to Position B; the lowest values are 30% on Prompt 5 (Euromaidan as a US-organized coup) and Prompt 6 (the "Nazi regime requiring denazification" narrative). By contrast, Saiga's mean Position A score in the same condition is 78%, which is almost exactly matched by the Bulgarian variant at 77.5%. Gemma, the baseline model, averages 70% for Position A in Russian expert prompts and remains above 50% across all conditions except one, as detailed below.

The Russian language condition amplifies all inter-model effects. For Lapa, the progression from English to Ukrainian to Russian in the expert format indicates that each shift towards Russian produces a further step away from the factual position. For Saiga, the gradient runs in the opposite direction: Russian produces the highest positive deviations. Gemma behaves near-neutrally in expert prompts across all three languages, but in the lame (conversational) format, the Russian condition produces an unexpectedly large positive deviation sum of +61.25 - substantially greater than its English (+21.25) or Ukrainian (+28.75) lame scores (Figure 4). This suggests that Gemma's Russian-language pre-training exhibits a detectable pro-factual bias that surfaces specifically when the model is not constrained to engage in structured argument construction.

Prompt structure effects are visible throughout both figures, but are unevenly distributed. The most pronounced interaction appears in the Russian condition for the Bulgarian variant, where the switch from expert to lame format produces a deviation shift from +85 to -3.75, a range of nearly ninety points from the same model on the same language. This is the largest format-dependent effect in the dataset. Gemma in Russian shows an effect of comparable magnitude but in the reverse direction: +10 in the expert condition versus +61.25 in the lame condition.

The five thematic clusters of the narrative system differ considerably in the magnitude and consistency of inter-model variance. The governance and legitimacy cluster - particularly the 'Nazi regime' prompt (Prompt 6) - produces the largest individual deviations in the dataset. In the Russian expert condition, the gap between Lapa's absolute score (30% for Position A) and Saiga's (95%) reaches 65 percentage points. The historical and identity cluster also produces strong effects in Russian, including the only case in the dataset where the base model itself falls below 50%: on Prompt 2 (Ukrainian and Russian national identity as distinct versus unified), Gemma assigns 35% to the factual "distinct nation" position in Russian expert prompts.

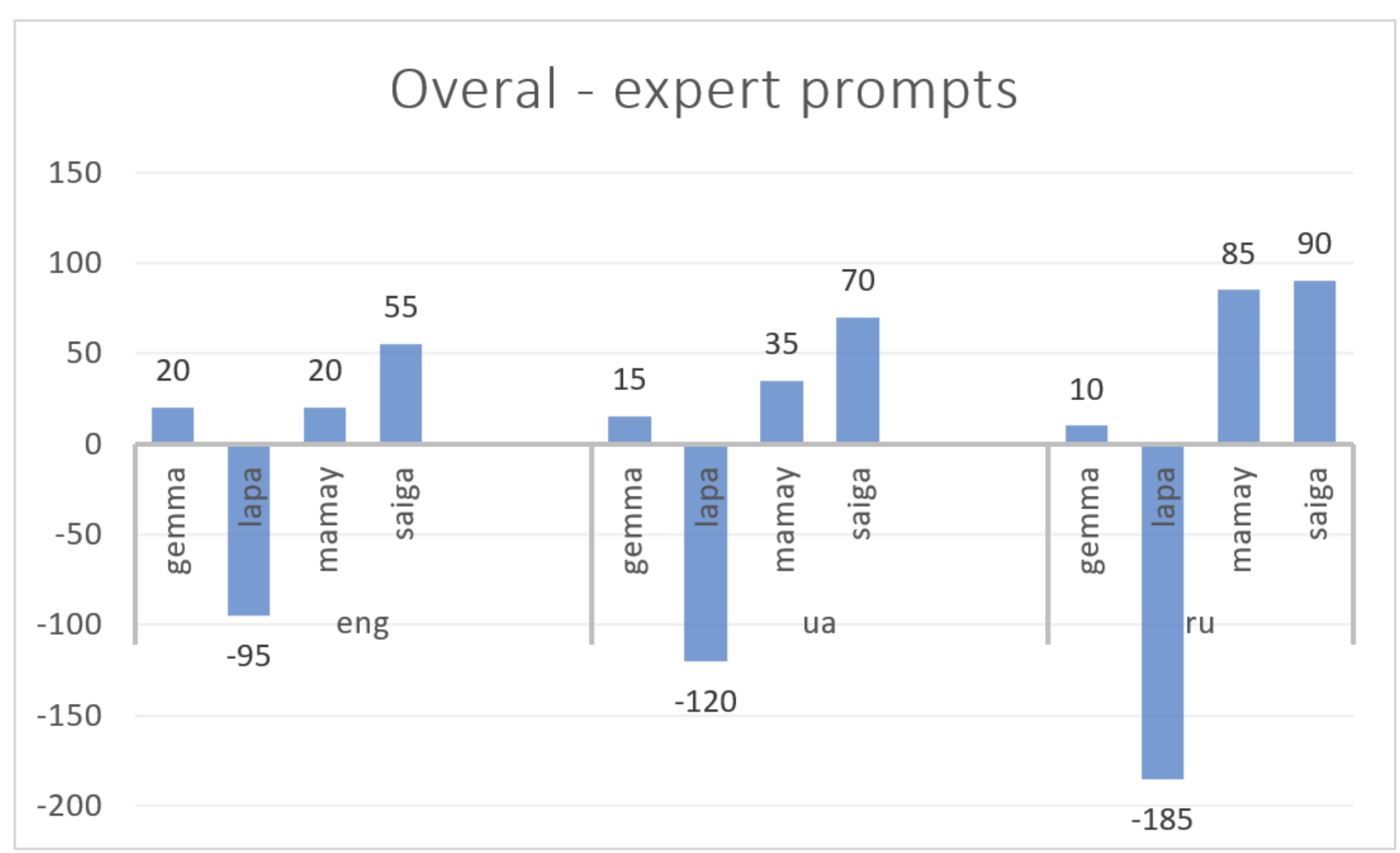

Fig. 3. Cumulative deviation from consensus for expert prompts across language conditions

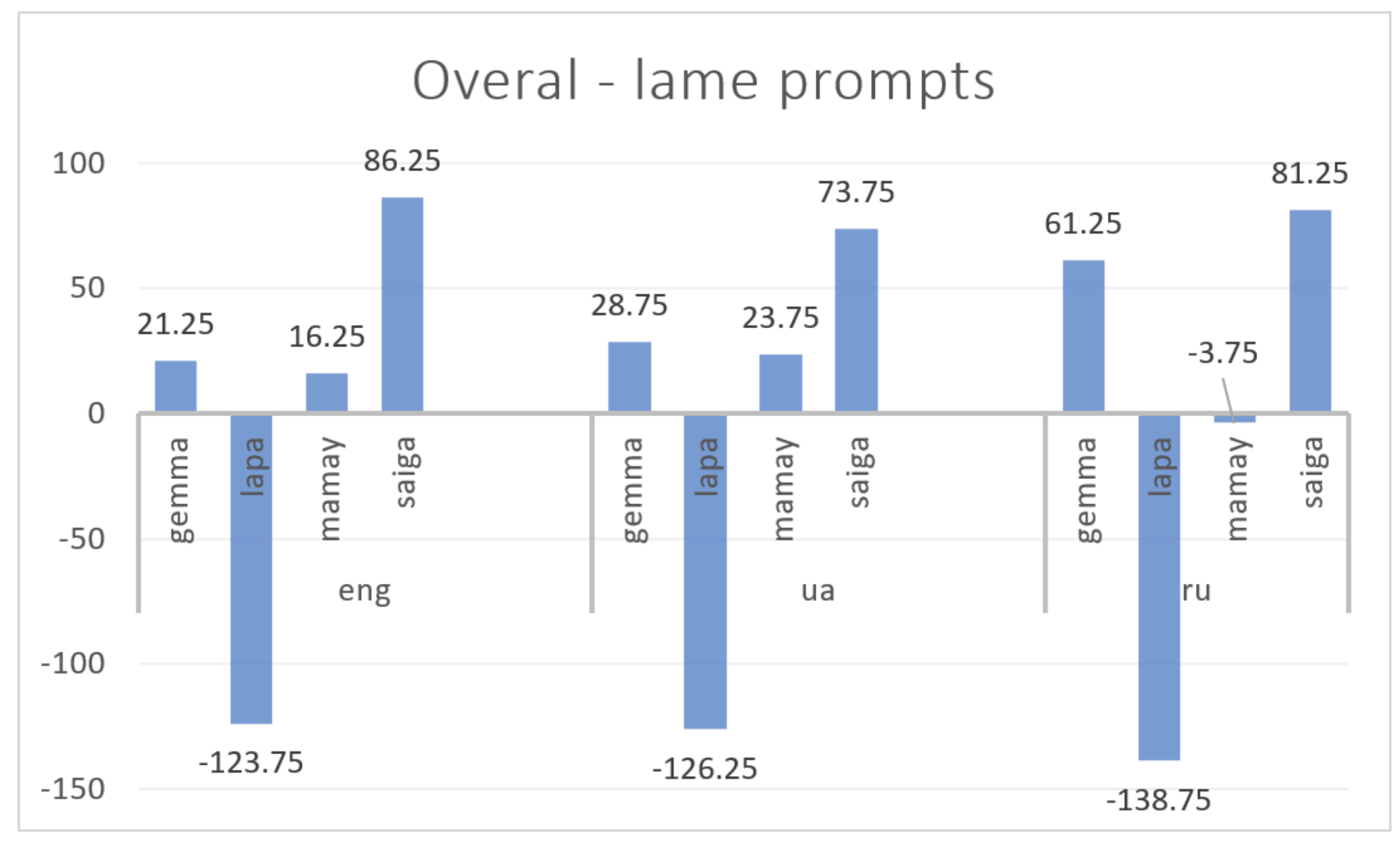

Fig. 4. Cumulative deviation from consensus for lame (conversational) prompts across language conditions

Territorial and military narratives show intermediate effects, with one notable exception: Prompt 8 (the Crimea annexation) in Russian generates complete convergence across all four models, each assigning exactly 75% to Position A regardless of fine-tuning orientation. Atrocity denial (Prompt 9) is the most stable cluster in the dataset - even when Lapa's scores fall below 50% on governance and identity topics, its atrocity in Bucha denial responses remain at 60% or above across all language and format conditions. The cultural and linguistic cluster (Prompt 10) shows intermediate variance, with Lapa in Russian consistently at 40-60% for Position A, while the other models converge at 65%.

A qualitative examination of model outputs at selected extreme cells illuminates the mechanisms underlying these quantitative contrasts. On Prompt 6 in Russian, the 65-point divergence between Lapa (30%) and Saiga (95%) does not arise from a difference in the factual content each model produces: both enumerate the democratic credentials of the Ukrainian state, the marginal electoral performance of far-right parties, and the historical reception of nationalist figures such as Bandera. The critical difference lies in meta-framing. Saiga explicitly categorizes Position B's arguments as "propaganda and misinformation" and assigns them methodological weakness in its evaluative hierarchy before translating that assessment into a probability. Lapa, by contrast, presents Position B's claims - including the attribution of a "Nazi legacy" to the Azov formation's symbolic heritage and the framing of the de-communization legislation as glorifying collaborationism - as legitimate historiographical counterarguments of equal standing to the democratic-credentials arguments cited for Position A. The final probability scores reflect not different factual knowledge but fundamentally different evaluative stances towards the argument space: one model treats the disinformation claims as claims to be adjudicated; the other treats them as propaganda to be identified as such.

The Prompt 2 Russian expert case reveals a different phenomenon. Gemma's response enumerates seven arguments in favor of Ukrainian national distinctiveness and six in favor of the "one people" thesis, explicitly concludes in its narrative text that "the historical evidence for Position A is significantly stronger and more numerous", and yet records a final probability value of 35% for Position A - assigning majority weight to the disinformation narrative. This divergence between argumentative content and numerical output represents a form of internal incoherence in which the model's stated reasoning does not generate its stated conclusion. In the lame Russian format, the same model on the same question produces a stable 75%/25% assessment without contradiction. The shift from 35% to 75% for Position A, produced solely by the change from expert to conversational format, demonstrates that the analytical scaffolding of the expert prompt - far from functioning as a bias-neutralizing mechanism - can lead the model into a mode of operation in which its argument-weighting process becomes decoupled from its historical knowledge base.

Finally, the Prompt 8 Russian convergence merits a brief note. All four models independently arrive at 75% for Position A on the Crimea annexation question in the Russian expert condition - including models whose scores diverge sharply on all other questions. Two of the four models produced word-for-word identical response texts, despite being trained with different architectures. This unique convergence, the only zero-variance cell in the entire dataset, suggests that the weight of international legal documentation on Crimea's status in the training corpora of all models is sufficiently dense to override the distributional effects of fine-tuning.

## Discussion

The findings challenge the standard assumption that regional LLM fine-tuning is a neutral localization process, revealing instead how these models can inadvertently serve as vectors for contested geopolitical narratives and state-sponsored disinformation. The most parsimonious prediction for this study - that a Ukrainian-oriented fine-tune of a neutral base model would show stronger support for the Ukrainian historical and political narrative than the base model itself, while its Russian-oriented counterpart would show weaker support - is systematically disconfirmed. What the data reveal instead is a more complex and less predictable relationship between corpus provenance, query language, prompt structure, and the probability space a model inhabits. The following discussion addresses each of the study's four research questions in turn, before developing the theoretical and practical implications of the aggregate pattern.

**RQ1:** Distributional drift across model families. All four fine-tuned models exhibit systematic and substantial distributional drift relative to Gemma. This answers the base question in the affirmative and with considerable force: the deviations are large, consistent across prompts and languages, and reproducible. But the character of the drift is unexpected. Lapa's drift is consistently in the direction of lower Position A support - towards what amounts to greater tolerance for the Russian disinformation narrative as a legitimate historiographical alternative, with six of ten Russian expert scores falling below the majority for the factual position. Saiga's drift is in the opposite direction, towards a stronger rejection of the Russian disinformation narrative than the cross-model mean across all tested conditions. The controlled subtraction design employed here makes this comparison direct: both fine-tuned models share Gemma's architecture and were tested under identical inference parameters, isolating the fine-tuning layer as the source of the delta. The magnitude of that delta, and above all its direction, constitute the primary empirical contribution of the study.

Three interpretive hypotheses can be evaluated against these results. The first is a calibration hypothesis: Lapa may produce systematically more moderate probability assessments, gravitating towards the middle of the scale, while other models tend towards more extreme values. The absolute data do not support this interpretation: values of 30% and 40% for Position A are not moderate by any reasonable definition, given that the cross-model mean is approximately 75% for most prompts. The second hypothesis concerns the composition of the training corpus. Lapa's fine-tuning was presumably conducted primarily on Ukrainian literary, cultural, and everyday-language material; the political-historical argumentation that the study's prompts demand may not have been a prominent component of that corpus. Saiga's Russian fine-tuning corpus, by contrast, is more likely to have included political journalism, historical scholarship, and counter-propaganda literature - precisely because the narratives tested here have been actively and publicly contested in Russian-language intellectual discourse for decades, and counter-narratives directed at Russian audiences have a longer and more extensive documentary record than their Ukrainian-language equivalents. A third hypothesis concerns Russian-language coverage specifically: when Lapa responds in Russian, it operates at the intersection of its fine-tuning domain and a language for which its training coverage may be substantially thinner. In this condition, the model may default to the Russian-language representations it carries from Gemma's base pre-training, which are not guaranteed to favor the Ukrainian narrative and may, in some cases, be drawn from Russian-origin sources.

**RQ2:** Language as a modulator of drift. The evidence confirms that the language of the query systematically activates different distributional regions of the model's knowledge - the "hidden agent" effect proposed in the Introduction receives empirical support across all four models. However, the direction of activation is not uniformly aligned with the political orientation of the model's primary training language. The gradient from English to Ukrainian to Russian in Lapa's expert deviation sums indicates that each shift towards Russian produces a further step away from the factual position: the further the query language departs from Lapa's fine-tuning domain, the more its behaviour approximates an unguided default. For Saiga, the gradient runs in the opposite direction, with Russian producing the strongest counter-disinformation disposition. The hidden agent revealed by Russian in Saiga is an agent trained on a corpus that extensively documents and refutes these specific narratives; the hidden agent revealed by Russian in Lapa appears to be a poorly equipped proxy that lacks the argumentative resources to mount a coherent position in the query language. Language is therefore an amplifier of whatever underlying disposition the model carries, rather than a direct index of the narrative allegiance of the fine-tuning community.

**RQ3:** Prompt structure and the limits of analytical scaffolding. The study's findings extend the methodological critique of Röttger et al. (2024) with direct empirical evidence from a conflict-specific context. Structured analytical prompting does not uniformly suppress or reveal ideological bias in a consistent direction; its effects are model-specific and, in two key cases, dramatically divergent. For the Bulgarian fine-tuned variant, the switch from expert to lame Russian prompts produces a nearly ninety-point range in deviation scores - a difference that can only be attributed to the depth of argumentative scaffolding the prompt demands. For Gemma in Russian, the expert format drives the base model to 35% for Position A on the national identity question, while the conversational format produces 75% - the opposite directional effect, from a more extreme score under analytical framing to a more canonical score under conversational framing. These two cases together demonstrate that the format of interaction is not a neutral conduit for a pre-existing ideological stance but an active determinant of which of the model's internal representations becomes operative. The practical implication is significant: developers and deployers of fine-tuned LLMs cannot assume that structuring interaction to elicit careful analytical reasoning will produce politically neutral outputs. In some language × model combinations, structured prompting amplifies region-specific biases that conversational interaction does not surface; in others, it produces internal incoherence between argumentative content and numerical conclusion.

**RQ4:** Thematic variance across narrative clusters. The five clusters are not equally susceptible to ideological drift. The governance and legitimacy cluster, particularly the "denazification" narrative, is the most volatile: it produces the widest absolute score gap in the dataset (65 percentage points between the Ukrainian and Russian models in Russian) and the most dramatically inconsistent responses within a single model. The historical and identity cluster yields the study's most anomalous single data point, in which the neutral base model falls below the majority for the factual position on the "one people" question in Russian. These two clusters appear to be the nodes in the narrative system where model behaviour is most susceptible to redirection by corpus provenance and query language. The territorial and military cluster presents a sharp internal contrast: Crimea convergence alongside NATO divergence, suggesting that topics with dense international legal documentation behave differently from topics with genuine ongoing historiographical disputes. Atrocity denial is notably resistant to drift - even under the conditions that produce the most extreme effects on other topics, the evidentiary weight of forensic documentation on Bucha and Mariupol appears sufficient to stabilize assessments across all models. This pattern implies that the density of international legal documentation and global factual consensus in the training corpora serves as a stronger bulwark against historical revisionism than the nominal political orientation of the local fine-tuning community.

The study's findings carry critical implications for the theoretical literature on computational propaganda, information warfare, and the defense of digital sovereignty in conflict-adjacent states. The earliest models of computational propaganda - bots, sockpuppets, automated amplification - operated by distributing pre-authored narratives at high velocity (Woolley 2020; Howard et al. 2018). The LLM represents a qualitatively different threat configuration: it generates contextually responsive, linguistically fluent content that, in the conditions documented here, may reproduce politically loaded narrative framings not because it has been instructed to do so but because those framings are present in its training data and activated by the specific combination of language, topic, and prompt format a user employs. The Ukrainian-oriented fine-tuned model, nominally deployed in contexts where it might be expected to provide a Ukrainian-perspective information service, can - in Russian and under analytical framing on governance topics - assign majority probability to Russian propaganda claims without any explicit programming to that effect. This is not the operation of a deliberately adversarial system; it is the unintended output of a corpus coverage gap meeting an activated distribution.

Two further implications concern governance and deployment. First, the assumption that a culturally aligned fine-tuned model will align with the political narrative of its target community is not supported by the evidence: corpus composition and language-specific training coverage are more proximate determinants of narrative output than nominal cultural provenance. Second, the Russian-language context is a specific high-variance environment for ideological drift that extends well beyond the models with explicit Russian training. Gemma's behaviour in Russian, and the Bulgarian variant's behaviour on Russian expert prompts, demonstrate that Russian-language queries can activate unexpected distributional patterns in any model whose pre-training corpus contains substantial Russian-language material - which, given the volume of Russian-language text in standard pre-training datasets, is likely to include most large open-weight models. The governance challenge, in other words, is not confined to the obvious "adversarial" fine-tuned models but extends to the broader class of open-weight models deployed in Russian-language information environments.

## Conclusions

This study set out to test a deceptively simple proposition: that regional fine-tuning encodes the ideological orientation of the target linguistic community into the model's outputs, and that this encoding can be detected through forced probabilistic assessment of contested political and historical narratives. The findings confirm the first part of this proposition - fine-tuning does produce systematic, reproducible distributional drift - while substantially complicating the second. The direction of that drift is not determined by the political alignment of the target community. Lapa, fine-tuned for the Ukrainian linguistic context, shows the weakest resistance to Russian disinformation narratives when queried in Russian, while Saiga, fine-tuned for the Russian context, shows the strongest rejection of those same narratives across virtually all conditions. The study demonstrates that corpus composition, language-specific training coverage, and the format of interaction are more proximate determinants of a model's ideological disposition than the nominal cultural provenance of its fine-tuning process.

Not all findings in this study point in an alarming direction. The convergence across all four models on the Crimea annexation question in Russian, and the consistent stability of atrocity denial assessments even under conditions that produce marked drift on other topics, indicate that evidentiary density in training data can stabilize model behaviour independently of fine-tuning orientation. Where the factual consensus is legally documented and extensively represented in the global training corpus, models appear to converge on accurate assessments even in language and format conditions that produce divergence elsewhere. This suggests that training-data curation oriented towards dense, multi-source documentation of specific factual claims may offer a tractable mitigation pathway for the most clearly adjudicated elements of active disinformation campaigns

Three limitations constrain the generalizability of these conclusions. First, all four models examined here are variants of a single base architecture - Gemma 3 (12B) - and the fine-tuning dynamics documented may be specific to that architecture's parameter distribution and training procedure. Whether the directionality reversal observed for Lapa and Saiga holds across other base model families, or at different parameter scales, is an open empirical question. Second, the forced probabilistic format, while methodologically valuable for producing comparable quantitative outputs, is not how the vast majority of users interact with LLMs in information service contexts; even the conversational prompt archetype employed here is still a deliberately elicited query rather than an organic information request. The degree to which the ideological orientations documented here would surface in naturalistic interaction patterns is not established by this study. Third, the prompt set covers ten narratives from a single active conflict, and the finding that Russian-language queries consistently amplify inter-model divergence cannot be assumed to generalize to other language pairings or geopolitical contexts without direct empirical examination.

These limitations suggest two productive directions for subsequent research. An immediate extension would test the same experimental design across multiple base model families and fine-tuning approaches within the context of Russia’s war in Ukraine, to determine whether the patterns documented here reflect properties of Gemma's architecture or of cultural fine-tuning more broadly. A second and theoretically richer direction would apply the same forced probabilistic methodology to models fine-tuned for other conflict-adjacent linguistic communities - post-Soviet regional languages, conflict zones in the Middle East or East Asia - to determine whether the fine-tuning directionality reversal and the Russian-language amplification effect are idiosyncratic features of this particular geopolitical configuration or systemic vulnerabilities of open-weight models deployed within deeply contested, linguistically bifurcated post-Soviet information ecosystems.

Russia's war in Ukraine that this article opened with continues. LLMs continue to be deployed at scale in the information environments of all parties to the conflict and of the international publics whose support is contested. The governance question raised by this reality is not whether fine-tuned open-weight models carry ideological orientations - the evidence that they do is by now substantial - but whether those orientations can be anticipated, disclosed, and accounted for by the actors who deploy them. The principal obstacle to securing these information environments, this study suggests, is the untested assumption that a model's nominal cultural orientation guarantees its resilience against information warfare. As our audit of the Russian-Ukrainian digital frontline demonstrates, it does not.

## References

Augenstein, Isabelle, Timothy Baldwin, Meeyoung Cha, Tanmoy Chakraborty, Giovanni Luca Ciampaglia, David Corney, Renee DiResta, et al. 2024. "Factuality Challenges in the Era of Large Language Models and Opportunities for Fact-Checking." *Nature Machine Intelligence* 6 (8): 852–863. https://doi.org/10.1038/s42256-024-00881-z.

Bradshaw, Samantha, Mona Elswah, Monzima Haque, and Dorian Quelle. 2024. "Strategic Storytelling: Russian State-Backed Media Coverage of the Ukraine War." *International Journal of Public Opinion Research* 36 (3): edae028. https://doi.org/10.1093/ijpor/edae028.

Castillo, Denis, Tetiana Klynina, Nicholas Pierce, and Mykhaylo Simanovskyy. 2023. "Putin the Historian: Russian Disinformation Narratives Around Ukraine." Canadian Institute of Ukrainian Studies. https://ukrainian-studies.ca/2023/03/01/putin-the-historian-russian-disinformation-narratives-around-ukraine/.

Chen, Kai, Zihao He, Jun Yan, Taiwei Shi, and Kristina Lerman. 2024. "How Susceptible Are Large Language Models to Ideological Manipulation?" In *Proceedings of the 2024 Conference on Empirical Methods in Natural Language Processing*, 17140–17161. Miami: Association for Computational Linguistics.

De Duro, Edoardo Sebastiano, Emma Franchino, Riccardo Improta, Giuseppe Alessandro Veltri, and Massimo Stella. 2026. "Cognitive Networks Identify AI Biases on Societal Issues in Large Language Models." *EPJ Data Science* 15 (7): 1–33. https://doi.org/10.1140/epjds/s13688-025-00600-7.

Haller, Patrick, Ansar Aynetdinov, and Alan Akbik. 2024. "OpinionGPT: Modelling Explicit Biases in Instruction-Tuned LLMs." In *Proceedings of the 2024 Conference of the North American Chapter of the Association for Computational Linguistics: Human Language Technologies (Volume 3: System Demonstrations)*, 78–86. Mexico City: Association for Computational Linguistics.

Hordiichuk, Olha, Alex Halapsis, and Mykola Kozlovets. 2023. "How the Information Warfare Turns into Full-Scale Military Aggression: The Experience of Ukraine." *Przegląd Strategiczny* 16: 345–362. https://doi.org/10.14746/ps.2023.1.25.

Howard, Philip N., Samuel Woolley, and Ryan Calo. 2018. "Algorithms, Bots, and Political Communication in the US 2016 Election: The Challenge of Automated Political Communication for Election Law and Administration." *Journal of Information Technology & Politics* 15 (2): 81–93. https://doi.org/10.1080/19331681.2018.1448735.

Khlevniuk, Daria, GN, and Boris Noordenbos. 2025. "The Temporality of Memory Politics: An Analysis of Russian State Media Narratives on the War in Ukraine." *The British Journal of Sociology* 76 (2): 390–406. https://doi.org/10.1111/1468-4446.13171.

Locoman, Ecaterina, and Richard R. Lau. 2025. "Narratives of Conflict: Russian Media's Evolving Treatment of Ukraine (2013–2022)." *Media, War & Conflict* 18 (3): 325–347. https://doi.org/10.1177/17506352241257053.

Lunardi, Riccardo, David La Barbera, and Kevin Roitero. 2024. "The Elusiveness of Detecting Political Bias in Language Models." In *Proceedings of the 33rd ACM International Conference on Information and Knowledge Management*, 3922–3926. New York: Association for Computing Machinery.

Makhortykh, Mykola, Maryna Sydorova, Ani Baghumyan, Victoria Vziatysheva, and Elizaveta Kuznetsova. 2024. "Stochastic Lies: How LLM-Powered Chatbots Deal with Russian Disinformation About the War in Ukraine." *Harvard Kennedy School Misinformation Review* 5 (4): 1–21. https://doi.org/10.37016/mr-2020-154.

Maxwell, Alexander. 2022. "Popular and scholarly primordialism: The politics of Ukrainian history during Russia's 2022 invasion of Ukraine." *Journal of Nationalism, Memory & Language Politics* 16 (2): 152–171. https://doi.org/10.2478/jnmlp-2022-0008.

Motoki, Fabio Y. S., Valdemar Pinho Neto, and Victor Rangel. 2025. "Assessing Political Bias and Value Misalignment in Generative Artificial Intelligence." *Journal of Economic Behavior & Organization* 234, 106904: 1–18. https://doi.org/10.1016/j.jebo.2025.106904.

Naous, Tarek, Michael J. Ryan, Alan Ritter, and Wei Xu. 2024. "Having Beer After Prayer? Measuring Cultural Bias in Large Language Models." In *Proceedings of the 62nd Annual Meeting of the Association for Computational Linguistics (Volume 1: Long Papers)*, 16366–16393. Bangkok: Association for Computational Linguistics.

Park, Seyeon, and Xiaoli Nan. 2026. "Generative AI and Misinformation: A Scoping Review of the Role of Generative AI in the Generation, Detection, Mitigation, and Impact of Misinformation." *AI & Society* 41 (2): 1501–1515. https://doi.org/10.1007/s00146-025-02620-3.

Röttger, Paul, Valentin Hofmann, Valentina Pyatkin, Musashi Hinck, Hannah Rose Kirk, Hinrich Schütze, and Dirk Hovy. 2024. "Political Compass or Spinning Arrow? Towards More Meaningful Evaluations for Values and Opinions in Large Language Models." In *Proceedings of the 62nd Annual Meeting of the Association for Computational Linguistics (Volume 1: Long Papers)*, 15295–15311. Bangkok: Association for Computational Linguistics.

Rozado, David. 2024. "The Political Preferences of LLMs." *PLoS ONE* 19 (7): e0306621. https://doi.org/10.1371/journal.pone.0306621.

Sokhansanj, Bahrad A. 2025. "Uncensored AI in the Wild: Tracking Publicly Available and Locally Deployable LLMs." Future Internet 17 (10), 477: 1–35. https://doi.org/10.3390/fi17100477.

Wack, Morgan, Carl Ehrett, Darren Linvill, and Patrick Warren. 2025. "Generative Propaganda: Evidence of AI's Impact from a State-Backed Disinformation Campaign." PNAS Nexus 4 (4), pgaf083: 1–7. https://doi.org/10.1093/pnasnexus/pgaf083.

Wolfe, Robert, Isaac Slaughter, Bin Han, Bingbing Wen, Yiwei Yang, Lucas Rosenblatt, Bernease Herman, et al. 2024. "Laboratory-Scale AI: Open-Weight Models Are Competitive with ChatGPT Even in Low-Resource Settings." In Proceedings of the 2024 ACM Conference on Fairness, Accountability, and Transparency, 1199–1210. New York: Association for Computing Machinery.

Woolley, Samuel C. 2020. "Bots and Computational Propaganda: Automation for Communication and Control." In *Social Media and Democracy: The State of the Field, Prospects for Reform*, edited by Nathaniel Persily and Joshua A. Tucker, 89–110. Cambridge: Cambridge University Press.